\documentclass[prb,preprint]{revtex4}
\usepackage{graphicx}
\usepackage{epsfig}
\usepackage{amsmath}

\begin{document}

\title{The force, power and energy of the 100 meter sprint}

\author{O. Helene}
\affiliation{Instituto de F\'{\i}sica, Universidade de S\~{a}o Paulo, 
C.P. 66318, CEP 05315-970 S\~{a}o Paulo, SP, Brazil}
\author{M. T. Yamashita}
\affiliation{Instituto de F\'\i sica Te\'orica, UNESP - Univ Estadual Paulista, C.P. 70532-2, 
CEP 01156-970, S\~ao Paulo, SP, Brazil}

\begin{abstract}
At the 2008 Summer Olympics in Beijing, Usain Bolt broke the world record for 
the 100\,m sprint. Just one year later, at the 2009 World Championships in Athletics in Berlin 
he broke it again. A few months after Beijing, Eriksen et al. studied Bolt's 
performance and predicted that Bolt could have run about one-tenth 
of a second faster, which was confirmed in Berlin. In this paper we extend the 
analysis of Eriksen et al. to model Bolt's velocity time-dependence for the Beijing 2008 and Berlin 
2009 records. We deduce the maximum force, 
the maximum power, and the total mechanical energy produced by Bolt in both 
races. Surprisingly, we conclude that all of these values were smaller in 2009 
than in 2008.
\end{abstract}


\maketitle

\section{Introduction}

Eriksen, Kristiansen, Langangen, and Wehus recently analyzed Usain Bolt's 100\,m sprint at 
the 2008 Summer Olympics in Beijing and speculated what the world record would have been had he 
not slowed down in celebration near the end of his race.\cite{bolt} They analyzed two scenarios for the last 
two seconds -- the same acceleration as the second place runner (Richard Thompson) or 
Bolt's acceleration was 0.5\,m/s$^2$ greater than Thompson's. 
They found race times of $9.61\pm0.04$\,s and $9.55\pm0.04$\,s for the first and second scenarios, respectively. 
In comparison, Bolt's actual race time in Beijing was 9.69\,s. The 
predictions in Ref.~\onlinecite{bolt} were apparently confirmed at the 2009 World Championships 
in Athletics in Berlin (race time 9.58\,s), suggesting that Bolt had improved his performance. However, 
the runners in Berlin benefited from a 0.9\,m/s tailwind that, as we will show, cannot be neglected.

In this paper we analyze Bolt's records in Beijing and Berlin using a function proposed 
by Tibshirani\cite{tib,wagner} to describe the time dependence of the velocity in both races. 
This paper is organized as follows. As a first step, we fit an empirical function, 
$v(t)$, to the data available from the International Association of Athletics 
Federation.\cite{iaaf} From this fit we can deduce the maximum acceleration, 
the maximum power, and the total mechanical energy produced by the runner. We then 
compare Bolt's performances in Berlin 2009 and Beijing 2008, concluding that all 
of these values were smaller in 2009 than in 2008. 

\section{Modeling the time dependence of the velocity}

In a seminal paper, Keller\cite{keller} analyzed the problem faced by runners -- how should they 
vary their velocity to minimize the time to run a certain distance? 
In Ref.~\onlinecite{keller} the resistance of the air was assumed 
to be proportional to the velocity and, for the 100\,m sprint, it was assumed that 
the best strategy is to apply as much constant force, $F$, as possible. 
The equation of motion was written as
\begin{equation}
m\frac{dv}{dt}=F-\alpha v,
\label{keller}
\end{equation}
where $v$ is the speed of the runner. The solution of Eq.~\eqref{keller} is $v(t)=(F/\alpha)(1-e^{-\alpha t/m})$. 
Although this solution seems to answer the proposed question, it has some limitations. One problem is that 
the air flow is turbulent, not laminar, and thus the drag force is proportional to $v^2$, rather than $v$. 
Another problem is that runners must produce energy (and thus a force) to replace the energy expended 
by the vertical movement of the runners' center of mass and the loss of energy due to the movement of 
their legs and feet. (See Ref.~\onlinecite{pfau} for an analysis of the energy cost 
of the vertical movement in a horse race.) Also, the runners' velocity declines slightly close to the 
end of the 100\,m race, which is not taken into account in Eq.~(\ref{keller}).

Tibshirani\cite{tib,wagner} proposed that a runner's force decreases with time according to 
$f(t)=F-\beta t$, so that the solution of Newton's second law is
\begin{equation}
v(t)=\left(\frac{F}{\alpha}+\frac{m\beta}{\alpha^2}\right)(1-e^{-\alpha t/m})-\frac{\beta t}{\alpha}.
\label{v}
\end{equation}
This solution solves some of the limitations discussed in the previous paragraph, 
but it is still incomplete. As we know, the air resistance is proportional to $v^2$, but 
in the above models the air resistance was supposed to be proportional only to $v$, 
probably to simplify the dynamical equations such that an analytical solution could 
be obtained.

The problem of determining the best form of the function to describe the time dependence of the velocity of a runner in the 100\,m 
sprint was addressed in Refs.~\onlinecite{summers,mureika1}, but is still open. 
We fitted Bolt's velocity in Beijing 2008 and Berlin 2009 using an equation with the 
same form as Eq.~(\ref{v}), but with three parameters to be fitted to his position as a function of time. 
We assume that the velocity 
is given by
\begin{align}
\label{vf}
v(t)&=a(1-e^{-ct})-bt, \\
\noalign{\noindent and the corresponding position is}
x(t)&=at-\frac{bt^2}{2}-\frac{a}{c}(1-e^{-ct}).
\label{xf}
\end{align}

\section{Experimental data and results}

The parameters $a$, $b$, and $c$ in Eq.~(\ref{xf}) were fitted using the least-squares method 
constrained such that $x(t_f)=100$\,m, where $t_f$ is the running time minus the `reaction time' 
(the elapsed time between the sound of the starting pistol and the initiation of the run). 
The split times for each 10\,m of Bolt's race are reproduced 
in Table~\ref{tab1}. In Table~\ref{tab1} we also show the split times calculated using the 
parameters given in Table~\ref{tab2}.

The data variances were estimated as follows. First, we assumed that the uncertainties 
of the split times are all equal. Then, we estimated the variances by the sum of the squares 
of the differences between the observed split times and the calculated values divided by the 
number of degrees of freedom (the number of data points minus the number of fitted parameters), 
$\nu$, considering both the Beijing and Berlin data:
\begin{equation}
\sigma^2=\frac{1}{\nu}\left(\sum_i^{\rm Beijing\;data}(t_i-t_i^{\rm fitted})^2+
\sum_i^{\rm Berlin\;data}(t_i-t_i^{\rm fitted})^2\right)
\end{equation}
The obtained result was  $\sigma=0.02$\,s. Then, we generated 100 sets of split times by summing 
normally distributed random numbers centered at zero with standard deviation equal to 0.02\,s to 
the observed split times.

For each simulation we fitted new values of the parameters. Thus, we obtained 100 values for each 
parameter, both to the Beijing and Berlin competitions. The uncertainties of the fitted parameters 
were estimated by the standard deviation of the 100 fitted parameters. The uncertainties calculated by 
this procedure reflect the uncertainties of the measured split times, the rounding of the published 
data, and the inadequacy of the model function.

Bolt's velocities in Beijing and Berlin are shown in Fig.~\ref{fig1}. In Berlin we 
observe a smaller initial acceleration and a smaller velocity decrease at the end of the race 
compared to his performance in Beijing. The mechanical power developed by Bolt is given by
\begin{equation}
P(t)=m\frac{dv}{dt}v+\kappa(v-v_{\rm wind})^2v+200{\rm W},
\label{pf}
\end{equation}
where the parameter $\kappa=C_d\rho A/2$ corresponds to the dissipated power due to drag and 
$v_{\rm wind}$ is the tailwind speed, which was zero in Beijing and equal to 0.9\,m/s in Berlin. For $C_d=0.5$, a typical value 
of the drag coefficient,\cite{keller,mureika1} $\rho=1.2$\,kg/m$^3$ for the air density, and a cross section of 
$A\approx1$\,m$^2$, we obtain $\kappa=0.3$\,W/(m/s)$^3$. A constant power of 200\,W is estimated for the power expended due 
to the vertical movement of the Bolt's center of mass.\cite{jump} We used 
$m=86$\,kg for his mass.\cite{charles} The mechanical power generated by Bolt is shown in Fig.~\ref{fig2}. We see that the maximum power 
generated by Bolt in Berlin is smaller than the power he generated in Beijing.

Table~\ref{tab3} shows Bolt's maximum acceleration and power at Beijing 2008 and at 
Berlin 2009 calculated using the fitted function $v(t)$. We 
also show the total mechanical energy produced in both competitions, calculated from the time integral 
of $P(t)$. Within the limitations of the model, Bolt's performance in Beijing (maximum force and power and the 
total energy) is higher than in Berlin. Also the predictions in Ref.~\onlinecite{bolt} seem 
to be vindicated.

\section{Discussion}

From Eq.~(\ref{vf}) we see that the velocity would become negative at times after the end of the race: $t=2.5$ and 6.5 
minutes, for Beijing and Berlin, respectively. Thus, the model is applicable to 
competitions of short duration, as can be observed by comparing the differences between 
the measured and fitted split times. The differences are very small and even in 
the worst cases (10\,m and 90\,m in the Beijing Olympic Games) the corresponding differences in the actual 
and calculated distances are not greater than 0.6\,m. The quality of the model function can also be 
observed by comparing the reported maximum velocity of 12.27\,m/s at 65.0\,m in Berlin with the result 
obtained from the fit equal to 11.92\,m/s at 57.6\,m.

An important question related to short races is the effect of the air resistance on the 
running time.\cite{pfau,linthorn,dapena,smith,quinn,mureika,mcfarland} 
In our model the power produced by the runner [see Eq.~(\ref{pf})] is used to accelerate the runner's 
body and to overcome air resistance (there also is a constant power of 200\,W, a kind of ``parasitic 
power loss'' dominated by the up-down movement of the runner's center of mass). If the runner is assisted by a 
tailwind, we assume that all the power saved to overcome air resistance will be used to accelerate the 
runner's body. Given this assumption, we ask what would the Beijing record have been for a 0.9\,m/s tailwind 
(the wind measurement on the day of the Berlin race)? We use Eq.~(\ref{pf}) and the same initial 
acceleration for Bolt in Beijing to deduce a new velocity profile in Berlin. We find that the record at 
the Beijing Olympic Games would have been reduced by 0.16\,s to 9.53\,s. Also, without the tailwind 
in Berlin and with the same power profile developed by the runner in this race, the time would increase 
by 0.16\,s.

The currently accepted theoretical and experimental results for the wind effect on 100\,m sprint 
times,\cite{pfau,linthorn,dapena,smith,quinn,mureika,mcfarland} gives a typical reduction of about 0.05\,s 
in the race time for a tailwind of 1\,m/s. A possible cause of our overestimation of the wind effect 
is the value of the drag coefficient $\kappa$ that we used. A smaller value of $\kappa$ implies that the 
wind effect is less important and, thus, the power saved in the presence of a tailwind would imply 
a smaller gain of velocity and, consequently, time. For instance, if we used $\kappa=0.1$\,W/(m/s)$^3$ the 
time reduction would be $\approx 0.06$\,s in the 100\,m sprint for a 0.9\,m/s tailwind -- 
close to the accepted value. Nevertheless, we chose $\kappa=0.3$\,W/(m/s)$^3$ (see the paragraph after 
Eq.~(\ref{pf})), which agrees with the recommended $\kappa$ values\cite{mureika1,smith,quinn} to estimate 
the energy and power produced by runners in the 100\,m sprint. 

Another possible origin for the discrepancy between our result for the wind effect and the literature is the 
force applied by the runner. We considered that the energy and the ground force applied by the 
runner can be deduced from the fitted position position versus time function. However, some workers prefer 
to formulate a model based on the applied forces (see, for instance, Refs.~\onlinecite{mureika1, linthorn,
dapena,quinn}). These authors usually obtain better agreement for the effects of wind as well as altitude 
in short races. 

\begin{acknowledgements}
MTY thanks FAPESP and CNPq for partial support. The authors thank P. Gouffon for a helpful discussion.
\end{acknowledgements}

\newpage
\section*{Tables}

\begin{table}[h!]
\centering
\begin{tabular}{|l|l|l|l|l|} \hline
& \multicolumn{4}{c|}{Split times\,(s)} \\ \hline
Distances\,(m) & Beijing 2008 & Fitted & Berlin 2009 & Fitted\\
\hline
10 & 1.85 & 1.91 & 1.89 & 1.91\\
\hline
20 & 2.87 & 2.88 & 2.88 & 2.89\\
\hline
30 & 3.78 & 3.77 & 3.78 & 3.78\\
\hline
40 & 4.65 & 4.63 & 4.64 & 4.63\\
\hline
50 & 5.50 & 5.47 & 5.47 & 5.46\\
\hline
60 & 6.32 & 6.31 & 6.29 & 6.29\\
\hline
70 & 7.14 & 7.15 & 7.10 & 7.11\\
\hline
80 & 7.96 & 7.95 & 7.92 & 7.93\\
\hline
90 & 8.79 & 8.84 & 8.75 & 8.76\\
\hline
100 & 9.69 & 9.69 & 9.58 & 9.58\\
\hline
\hline
Reaction time & 0.165 & & 0.146 &\\
\hline
\end{tabular}
\caption{\label{tab1}Bolt's split times at 10\,m intervals in Beijing 2008 and Berlin 2009. The estimated standard 
deviations of the data are 0.02\,s (see text). The last line shows the reaction times, which were subtracted from 
the split times before the fit. We also show the split times calculated using the fitted parameters shown in 
Table~\ref{tab2}. The data is from Ref.~\onlinecite{iaaf}.}
\end{table}

\begin{table}[h!]
\centering
\begin{tabular}{|l|l|l|} \hline
Parameters & 2008 & 2009 \\
\hline
$a$ & 12.49(3)\,{\rm m/s} & 12.43(3)\,{\rm m/s} \\
\hline
$b$ & 0.081(4)\,{\rm m/s}$^{-2}$ & 0.032(3)\,{\rm m/s}$^{-2}$ \\
\hline
$c$ & 0.814(4)\,{\rm s}$^{-1}$ & 0.783(3)\,{\rm s}$^{-1}$ \\
\hline
\end{tabular}
\caption{\label{tab2}Fitted parameters of Eqs.~(\ref{vf}) and (\ref{xf}). The estimation of the parameter uncertainties 
is described in the text.}
\end{table}

\begin{table}[h!]
\centering
\begin{tabular}{|l|l|l|} \hline
& Beijing 2008 & Berlin 2009 \\
\hline
Maximum acceleration & 10.09(3)\,{\rm m/s}$^2$ & 9.70(3)\,{\rm m/s}$^2$ \\
\hline
Maximum power & 2934(3)\,{\rm W} & 2827(3)\,{\rm W} \\
\hline
Total energy & 11611(7)\,{\rm J} & 11531(6)\,{\rm J} \\
\hline
\end{tabular}
\caption{\label{tab3}Maximum acceleration, maximum power, and total mechanical energy produced by Bolt in Beijing and Berlin. 
The standard deviations were calculated using the parameter uncertainties given in Table~\ref{tab2}.}
\end{table}

\newpage
\section*{Figure captions}

\begin{figure}[h!]
\caption[dummy0]{Fitted $v(t)$ of Usain Bolt in Beijing 2008 (solid line) and in Berlin 2009 (dashed line).} 
\label{fig1}
\end{figure}

\begin{figure}[h!]
\caption[dummy0]{Total mechanical power of Usain Bolt in Beijing 2008 (solid line) and in Berlin 2009 
(dashed line).} 
\label{fig2}
\end{figure}


\begin{thebibliography}{16}

\bibitem{bolt} H. K. Eriksen, J. R. Kristiansen, \O. Langangen, and I. K. Wehus, 
``How fast could Usain Bolt have run? A dynamical study,'' Am. J. Phys. {\bf 77}, 
224--228 (2009).

\bibitem{tib} R. Tibshirani, ``Who is the fastest man in the world?,'' Am. Stat. {\bf 51}, 106--111 (1997).

\bibitem{wagner} G. Wagner ``The 100-meter dash: Theory and experiment,'' Phys. Teach. {\bf 36}, 144--146 (1998).

\bibitem{iaaf} International Association of Athletics Federations, \url{<www.iaaf.org>}.

\bibitem{keller} J. B. Keller, ``Theory of competitive running,'' Phys. Today {\bf 26} (9), 43--47 (1973).

\bibitem{pfau} T. Pfau, A. Spence, S. Starke, M. Ferrari, and A. Wilson, ``Modern riding style improves horse 
racing times,'' Science {\bf 325}, 289 (2009).

\bibitem{summers} R. L. Summers, ``Physiology and biophysics of the 100-m sprint,'' 
News Physiol. Sci. {\bf 12}, 131--136 (1997).

\bibitem{mureika1} J. R. Mureika, ``A realistic quasi-physical model of the 100\,m dash,'' 
Can. J. Phys. {\bf 79}, 697--713 (2001).

\bibitem{jump} O. Helene and M. T. Yamashita, ``A unified model for the long and high jump,'' 
Am. J. Phys. {\bf 73}, 906--908 (2005).

\bibitem{charles} J. D. Charles and A. Bejan, ``The evolution of speed, size and shape in modern athetics,'' 
J. Exp. Biol. {\bf 212}, 2419--2425 (2009).

\bibitem{linthorn} N. P. Linthorne, ``The effect of wind on 100\,m sprint times,'' J. Appl. Biomech. {\bf 10}, 110--131 (1994).

\bibitem{dapena} J. Dapena and M. E. Feltner, ``Effects of wind and altitude on the times of 100 meter sprint races," 
Int. J. Sports Biomech. {\bf 3}, 6--39 (1987).

\bibitem{smith} A. J. Ward-Smith, ``New insights into the effect of wind assistance on sprinting performance,'' 
J. Sports Sci. {\bf 17}, 325--334 (1999).

\bibitem{quinn} M. D. Quinn, ``The effects of wind and altitude in the 200\,m sprint,'' J. Appl. Biomech. {\bf 19}, 49--59 (2003).

\bibitem{mureika} J. R. Mureika, ``Modeling wind and altitude effects in the 200\,m sprint,'' Can. J. Phys. {\bf 81}, 895--910 (2003).

\bibitem{mcfarland} E. McFarland, ``How Olympic records depend on location,'' Am. J. Phys. {\bf 54}, 513--519 (1986).

\end{thebibliography}
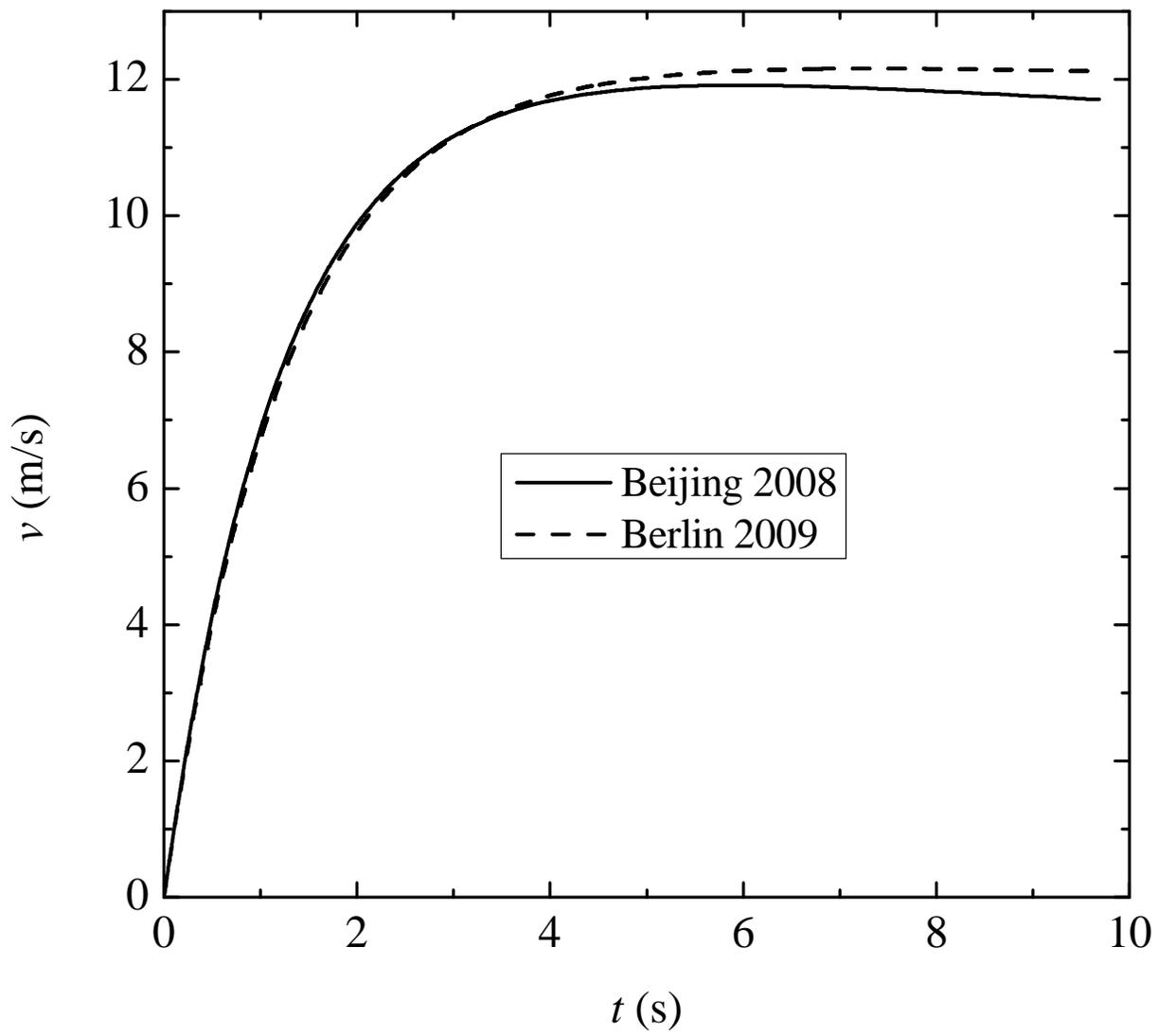
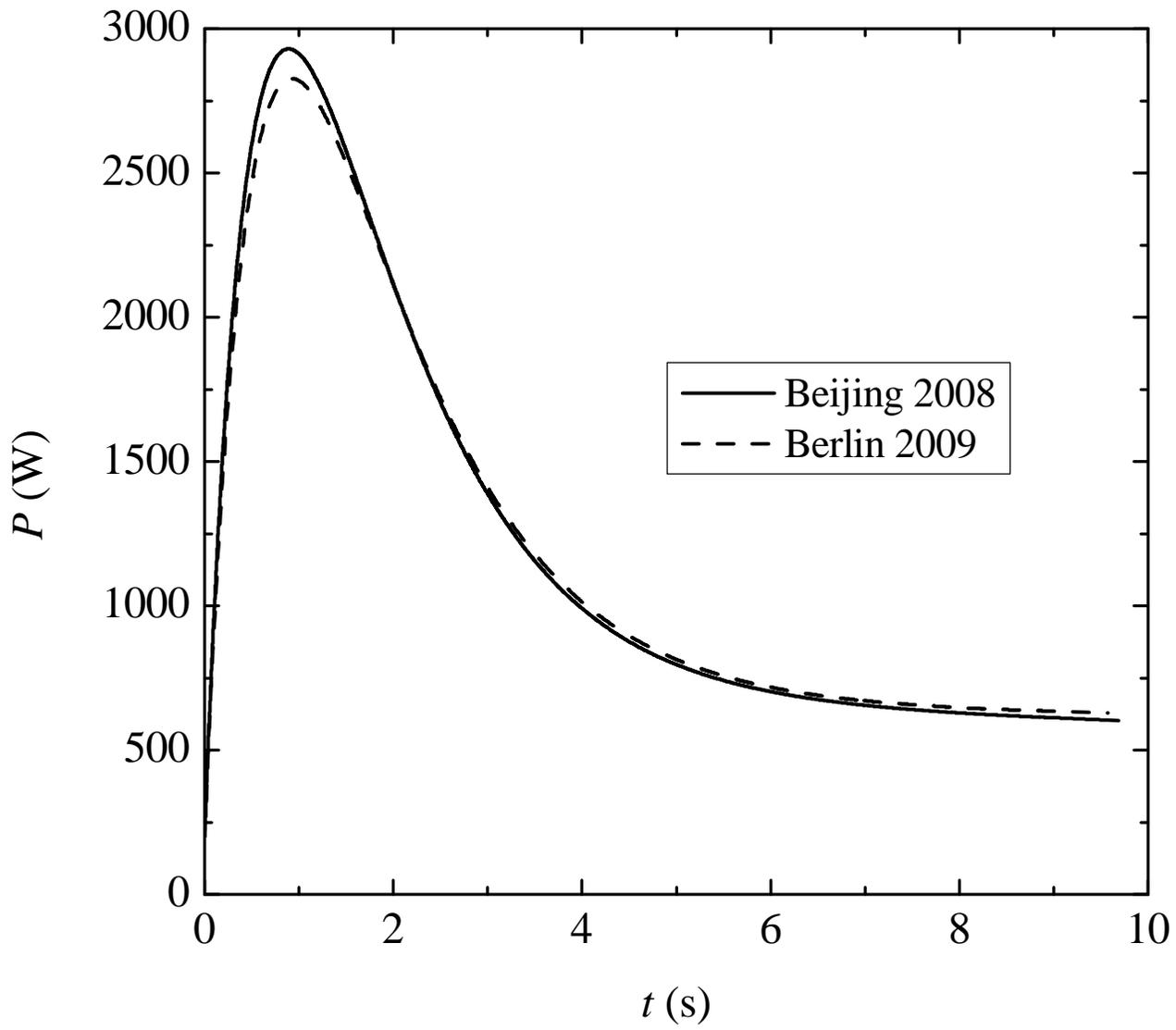
\end{document}